\documentclass[letterpaper, conference]{IEEEtran}
\usepackage{amsthm}

\usepackage{cite}
\usepackage[cmex10]{amsmath}
\usepackage{algorithm}
\usepackage{algorithmic}
\usepackage{breqn}

\usepackage[inline]{trackchanges}

\usepackage{flushend}

\usepackage{amsmath}
\usepackage{amsfonts}
\usepackage{amssymb}
\usepackage{mathtools}

\addeditor{swanand}
\addeditor{as}
\addeditor{brenden}
\addeditor{salim}

\usepackage{amsfonts, amssymb, amscd, xspace}

\usepackage{algorithm}
\usepackage{algorithmic}

\usepackage{color}
\usepackage{graphicx}
\usepackage{epstopdf}

\usepackage{xfrac}

\newtheorem{theorem}{Theorem}
\newtheorem{lemma}{Lemma}

\newtheorem{corollary}{Corollary}
\newtheorem{remark}{Remark}

\newtheorem{example}{Example}
\newtheorem{definition}{Definition}
\newtheorem{proposition}{Proposition}

\newcommand{\etal}{{\it et al.}}
\newcommand{\ie}{{\it i.e.}}
\newcommand{\eg}{{\it e.g.}}

\newcommand{\Xj}[1]{\boldsymbol{X}_{#1}} 

\newcommand{\Q}[2]{Q^{[#1, #2]}} 
\newcommand{\A}[2]{A^{[#1,#2]}} 
\newcommand{\Qb}[2]{\boldsymbol{Q}^{\left[#1, #2\right]}} 
\newcommand{\Ab}[2]{\boldsymbol{A}^{\left[#1,#2\right]}} 

\newcommand{\Hp}[1]{H\left(#1\right)} 
\newcommand{\Hc}[2]{H\left(#1 \mid #2\right)} 

\newcommand{\twomatrix}[2]{\begin{bmatrix} #1 \\ #2\end{bmatrix}}

\newcommand{\GF}[1]{\mathbb{F}_{#1}} 

\newcommand{\code}{\mathcal{C}} 
\newcommand{\dual}{\mathcal{C}^{\perp}} 

\newcommand{\vect}[1]{\mathbf{#1}} 
\newcommand{\cw}{\mathbf{c}} 

\newcommand{\supp}[1]{\textrm{Supp}\left(#1\right)} 

\newcommand{\rep}[2]{{R}_{#1}\left(#2\right)} 
\newcommand{\gam}[1]{\Gamma\left(#1\right)} 
\newcommand{\subspace}[1]{\left\langle #1 \right\rangle} 

\newcommand{\nkdq}{[n,k,d]_q} 
\newcommand{\cwi}{\vect{c}_i} 

\newcommand{\dm}[1]{d_{min}\left(#1\right)} 

\newcommand{\evec}[1]{e_{#1}}
\newcommand{\setA}{\mathcal{A}}

\newcommand\blfootnote[1]{%
  \begingroup
  \renewcommand\thefootnote{}\footnote{#1}%
  \addtocounter{footnote}{-1}%
  \endgroup
}

\renewcommand{\baselinestretch}{1.015}

\title{\LARGE \bf
On an Equivalence Between Single-Server PIR with Side Information and Locally Recoverable Codes}

\IEEEoverridecommandlockouts

\begin{document}



%
\author{Swanand Kadhe, Anoosheh Heidarzadeh, Alex Sprintson, and O. Ozan Koyluoglu}


%



\maketitle

\begin{abstract}
Private Information Retrieval (PIR) problem has recently attracted a significant interest in the information-theory community.
{In this problem,} a user wants to privately download one or more messages belonging to a database with copies stored on a single or multiple remote servers. 
{In the single server scenario, the user must have prior side information, \emph{i.e.}, a subset of messages unknown to the server, to be able to privately retrieve the required messages in an efficient way.}

{In the last decade, there has also been a significant interest in Locally Recoverable Codes (LRC), a class of storage codes in which each symbol can be recovered from a limited number of other symbols. More recently, there is an interest in \emph{cooperative} locally recoverable codes, \emph{i.e.}, codes in which multiple symbols can be recovered from a small set of other code symbols.}

In this paper, we {establish a} relationship between coding schemes for the single-server PIR problem and {LRCs}. 
In particular, we show the following results: (i) PIR schemes designed for retrieving a single message are ‘equivalent’ to classical LRCs; 
and (ii) PIR schemes for retrieving multiple messages are equivalent to cooperative LRCs. 
{These equivalence results allow us to recover upper bounds on the download rate for PIR-SI schemes, and to obtain a novel rate upper bound on cooperative LRCs.}
{We show results for both linear and non-linear codes.}

\end{abstract}

\blfootnote{
S. Kadhe and O. O. Koyluoglu are with the Department of Electrical Engineering and Computer Sciences at University of California Berkeley, USA; emails: \{swanand.kadhe, ozan.koyluoglu\}@berkeley.edu. 

A. Heidarzadeh and A. Sprintson are with the Department of Electrical and Computer Engineering at Texas A\&M University, USA; emails:\{anoosheh, spalex\}@tamu.edu.  

This work is supported in part by National Science Foundation grants CCF-1748585 and CNS-1748692. 

This material is based upon work supported while Alex Sprintson was serving at the National Science Foundation. Any opinion, findings, and conclusions or recommendations expressed in this material are those of the author and do not necessarily reflect the views of the National Science Foundation.
}

\section{Introduction}
\label{sec:intro}

The Private Information Retrieval (PIR) problem is one of the important problems in theoretical computer science \cite{chor1998private}. 
The setting of the problem includes a client that needs to retrieve a message belonging to a database with copies stored on a single or multiple remote servers. The message needs to be retrieved by satisfying the privacy condition, which prevents the server from identifying the index of the retrieved message. 
{The theoretical computer science community has primarily focused on the settings with small message sizes with the objective to minimize the total number of bits uploaded to and downloaded from the server (see~\cite{yekhanin2010private}).}

Starting with the seminal work of Sun and Jafar~\cite{Sun2017}, the multiple-server PIR problem has received a significant attention from the information and coding theory community with breakthrough results in the past few years (see, \eg, \cite{JafarPIR3new, tajeddine2017private1,Banawan2017,BU18}, and references therein). 
{The information-theoretic approach has focused on a practical setting with large message sizes with the goal to minimize the ratio of the total number of downloaded bits to the message size.}


Recently, {Kadhe et al. } \cite{KadheGHERS2017:Allerton,KadheGHERS2017}  considered the single-server PIR with Side Information (PIR-SI) problem, wherein the user knows a random subset of messages that is unknown to the server. 
It was shown that the side information enables the user to substantially reduce the download cost and still achieve information-theoretic privacy for the requested message. The multi-message extension of PIR-SI, which enables a user to privately download multiple messages from the server, is considered {by Heidarzadeh \etal}~\cite{HKGRS:2018} as well as Li and Gastpar~\cite{LG:2018}. 

It is well-known in the theoretical computer science community that there is a strong relationship between PIR schemes and a class of error-correcting codes called {\it locally decodable codes} (LDCs) (see, \eg, the surveys~\cite{yekhanin2010private,yekhanin2011locally}). 
LDCs allow one to locally decode an arbitrary message symbol from only a small subset of randomly chosen codeword symbols, even after a fraction of codeword symbols are corrupted by an adversary.

Continuing with this theme, in this paper, we show that single-server PIR-SI schemes are closely related to another class of codes with locality called {\it locally recoverable codes} (LRCs)~\cite{Gopalan:12}. LRCs are a class of erasure codes that enable one to recover an erased codeword symbol from only a small subset of other codeword symbols. 

In particular, in an LRC with block-length $n$ and locality $r$, every codeword symbol can be reconstructed from at most $r$ other codeword symbols~\cite{Gopalan:12}. {Rawat et al.}~\cite{Rawat:14:cooperative,Rawat:15:cooperative} extended the notion of local recovery to {\it cooperative local recovery}. Specifically, in an LRC with block-length $n$ and $(r,\ell)$-cooperative locality, every subset of $\ell$ codeword symbols can be reconstructed from at most $r$ other codeword symbols.

In this paper, we show that single-message PIR-SI schemes are related to LRCs, whereas multi-message PIR-SI schemes are related to cooperative LRCs. Detailed contributions are outlined in the following.

{\bf Our Contributions:} We focus out attention to the single- server PIR-SI problem in which a user wishes to download $D$ messages from a database of $K$ messages (over a finite field $\GF{q}$), stored on a single remote server. The user has a random subset of $M$ messages, {referred to as \emph{side information},} whose identities are unknown to the server. 

First, we focus on the scalar-linear case wherein the answer from the server is of the form $EX$, where $X = [X_1 \: \cdots \: X_K]^T\in\GF{q}^K$ denotes the set of messages, and $E$ is a $T\times K$ matrix with entries over $\GF{q}$. When the user wishes to protect only the identities of the requested messages, we show the following results:
\begin{itemize}
    \item Equivalence between single-message $(D=1)$ PIR {with Side Information} (SM-PIR-SI) schemes and LRCs:
    \begin{enumerate}
        \item {Any} solution $E$ to an SM-PIR-SI problem {is} a parity check matrix of an LRC with block-length $K$ and locality $M$ 
        (Theorem~\ref{thm:1-PIR-LRC}).
        \item Given a parity check matrix $H$ of an LRC with block-length $K$ and locality $M$,
        it is possible to construct an SM-PIR-SI scheme where $E$ is a column-permutation of $H$ (Theorem~\ref{thm:1-LRC-PIR}). 
    \end{enumerate}
    \item Equivalence between multi-message $(D\geq 2)$ PIR {with Side Information} (MM-PIR-SI) schemes and cooperative LRCs:
    \begin{enumerate}
        \item {Any} solution $E$ to a MM-PIR-SI problem {is} a parity check matrix of an LRC with block-length $K$ and $(M,D)$-cooperative locality (Theorem~\ref{thm:D-PIR-LRC}).
        \item Given a parity check matrix $H$ of an LRC with block-length $K$ and $(M,D)$-cooperative locality, it is possible to construct an MM-PIR-SI scheme where $E$ is a column-permutation of $H$ (Theorem~\ref{thm:D-LRC-PIR}). 
    \end{enumerate}
    \item 
    As corollaries to Theorems~\ref{thm:1-PIR-LRC} and~\ref{thm:D-PIR-LRC}, we derive upper bounds on the download rates for SM-PIR-SI problem (Corollary~\ref{cor:1-PIR-LRC}) and MM-PIR-SI problem (Corollary~\ref{cor:D-PIR-LRC}), respectively. In addition, as a corollary to Theorem~\ref{thm:D-LRC-PIR}, we derive a novel tight upper bound on the rate of a cooperative LRC for the regime $\ell > r$ (see Corollary~\ref{cor:rate-LRC} and Remark~\ref{rem:rate-LRC}).
\end{itemize}
Next, we consider the case when the user wants to protect both the identities of the requested messages and that of the side-information, referred to as $(W,S)$-PIR-SI.\footnote{Here, $W$ denotes the demand index set and $S$ denotes the side information index set. {We use the} term $(W,S)$-PIR-SI {to reflect} the fact that the user wants to protect $(W,S)$ jointly.} We show the following equivalence result:
\begin{itemize}
    \item Equivalence between $(W,S)$-PIR-SI schemes and maximum distance separable (MDS) codes\footnote{An MDS code can be considered as an LRC with locality $r = k$.}:
    \begin{enumerate}
        \item Any solution $E$ to a $(W,S)$-PIR-SI problem is a parity check matrix of an MDS code with block-length $K$ and dimension $M$ (Theorem~\ref{thm:PIR-MDS}).
        \item Given a parity check matrix $H$ of an MDS code with block-length $K$ and dimension $M$, it is possible to construct a $(W,S)$-PIR-SI scheme where $E = H$ (Theorem~\ref{thm:MDS-PIR}). 
    \end{enumerate}
\end{itemize}
Finally, we lift the restriction of scalar-linear solutions, and consider generic 
{(non-linear)} SM-PIR-SI schemes. We show the following equivalence result:
\begin{itemize}
    \item Equivalence between SM-PIR-SI schemes and LRCs with maximum possible size\footnote{It is possible to show that any LRC over $\GF{q}$ with block-length $n$ and locality $r$ can contain at most $q^{n - \lceil n/(r+1) \rceil}$ codewords (see Proposition~\ref{prop:max-size}). Any LRC with $q^{n - \lceil n/(r+1) \rceil}$ codewords is said to be {an LRC code} with maximum possible size.}:
    \begin{enumerate}
        \item Given a solution to an SM-PIR-SI problem, it is possible to construct an LRC with block-length $K$ and locality $M$ (Theorem~\ref{thm:1-PIR-LRC-gen}).
        \item Given an LRC with block-length $K$ and locality $M$ with the maximum possible size, it is possible to construct an SM-PIR-SI scheme (Theorem~\ref{thm:1-LRC-PIR-gen}). 
    \end{enumerate}
\end{itemize}

\section{Preliminaries}
\label{sec:preliminaries}

{\bf Notation:} For a positive integer $K$, denote $\{1,\dots,K\}$ by $[K]$. Let $\GF{q}$ denote the finite field of order 
$q$, where $q$ is a power of a prime. For a set $\{X_1,\dots,X_K\}$ and a subset $S\subset {[K]}$, let \mbox{$X_S = \{X_j : j\in S\}$}. 
For a positive integer $P$, let $\mathbf{1}_P$ and $\mathbf{0}_P$, respectively, denote the all-one and all-zero row vectors of length $P$.
Let $e_j$ be a unit vector of length $K$ such that its $j$-th entry is $1$ and the other entries are $0$. For a set $W = \{W_1,W_2,\ldots,W_D\}\subseteq[K]$, let 
$$
I_W = \begin{bmatrix}\evec{W_1}\\ \evec{W_2}\\ \vdots\\ \evec{W_D}\end{bmatrix}.
$$ 
For a $T\times K$ matrix $E\in\GF{q}^{T\times K}$, let $\langle E \rangle$ denote the row-space of $E$. 
For a subset $S\subset[K]$, let $E_S$ denote the $T\times|S|$ submatrix consisting of columns of $E$ indexed by $S$. For a vector $v$, let $\supp{v}$ denote the support of $\vect{v}$. For a subspace $\code\subset\GF{q}^K$, let ${\dual}$ be its dual subspace. 

\subsection{Single-Server PIR with Side Information}
\label{sec:PIR-basics}
We briefly overview the single-server PIR with side information problem~\cite{KadheGHERS2017:Allerton,HeidarzadehGKES:18} (see also~\cite{KadheGHERS2017}). Consider a server containing a database that consists of a set of $K$ messages $\boldsymbol{X} = [\Xj{1}\: \cdots \:\Xj{K}]^T$, with each message being independently and uniformly distributed over $\GF{q}$. 
A user is interested in {\it privately} downloading $D$ ($1\leq D\leq K$) messages $\Xj{W}$ from the server for some $W \subseteq [K]$, $|W|=D$. We refer to $W$ as the \emph{demand index set} and $\Xj{W}$ as the \emph{demand}. 
The user has the knowledge of a subset $\Xj{S}$ of the messages 
for some $S\subset [K]\setminus W$, $|S| = M$, $M\leq K-D$. 
We refer to $S$ as the \emph{side information index set} and $\Xj{S}$ 
as the \emph{side information}. 

Let $\boldsymbol{W}$ and $\boldsymbol{S}$ denote the random variables corresponding to the demand and side information index sets, respectively. 
{We assume that} the side information index set $\boldsymbol{S}$ is distributed uniformly over over all subsets of $[K]$ of size $M$, i.e., 
\begin{equation}
\label{eq:SideInfoDist}
p_{\boldsymbol{S}}(S) =
\left\{\begin{array}{ll}
\frac{1}{\binom{K}{M}}, & S\subset [K], |S|=M,\\
0, & \text{otherwise}.
\end{array}\right.
\end{equation}
Further, {we assume that} the demand index set $\boldsymbol{W}$ has the following conditional distribution given $S$: 
\begin{equation}\label{eq:WantSetDist}
p_{\boldsymbol{W}|\boldsymbol{S}}(W\mid S) = 
\left\{\begin{array}{ll}
\frac{1}{\binom{K-M}{D}}, & W\subseteq [K]\setminus S,|W|=D,\\
0, & \text{otherwise}.
\end{array}\right.	
\end{equation}
We assume that the {server does} not know the side information realization at the user and only knows the  {\it a priori} distributions 
$p_{\boldsymbol{S}}(S)$ and $p_{\boldsymbol{W}|\boldsymbol{S}}(W|S)$.

To download the set of messages $\Xj{W}$ given the side information $\Xj{S}$, the user sends a query $\Q{W}{S}$ 
to the server. The server responds to the query it receives with an answer $\A{W}{S}$ over $\GF{q}^T$. 
Let $\Qb{W}{S}$ and $\Ab{W}{S}$ be the corresponding random variables. 


\begin{definition}
\label{def:PIR}
[PIR-SI] Any scheme consisting of a query and an answer is referred to as the {\it PIR with side information (PIR-SI) scheme} if
the query and answer satisfy the following two conditions.
\begin{itemize}
\item[1.]  
$W$-\textbf{privacy}: The server cannot infer any information about the demand index set from the query it receives 
\ie, 
\begin{equation}
\label{eq:W-privacy}
I\left(\boldsymbol{W}; \Qb{\boldsymbol{W}}{\boldsymbol{S}}\right) = 0.
\end{equation}

\item[2.]  
$(W,S)$-\textbf{privacy}: The server cannot infer any information about the demand index set as well as the side information index set from the query it receives
\ie, 
\begin{equation}
\label{eq:WS-privacy}
I\left(\boldsymbol{W},\boldsymbol{S}; \Qb{\boldsymbol{W}}{\boldsymbol{S}}\right) = 0.
\end{equation} 

\item[3.] \textbf{Recoverability:} From the answer $A^{[W,S]}$ 
and the side information $X_S$, the user should be able to decode the desired set of messages $X_W$ for any $(W,S)$, i.e.,
\begin{equation}
\label{eq:decodability}
\Hc{\Xj{\boldsymbol{W}}}{\Ab{\boldsymbol{W}}{\boldsymbol{S}}, \Qb{\boldsymbol{W}}{\boldsymbol{S}},\Xj{\boldsymbol{S}},\boldsymbol{W},\boldsymbol{S}} = 0.
\end{equation} 
\end{itemize}
\end{definition}
We refer to the case of $D = 1$ as single-message PIR-SI, while the case of $D\geq 2$ as multi-message PIR-SI.

The \emph{rate} of a PIR-SI scheme is defined as the ratio of the message length ($\log q$ bits) to the total length of the answers (in bits) as follows:\footnote{We focus our attention to the download rate similar to~\cite{Sun2017}. This is because the download rate dominates the total communication rate when the message size is sufficiently large as compared to the size of a query.}
\begin{equation}
\label{eq:rate}
R = \frac{D\log q}{\Hp{\Ab{\boldsymbol{W}}{\boldsymbol{S}}}}.
\end{equation}
The \emph{capacity} of $W$-PIR-SI, denoted by $C_{W}$, is defined as the supremum  of rates over all $W$-PIR-SI schemes for a given $K$ and $M$.

\subsection{Locally Recoverable Codes}
\label{sec:LRC-basics}
Let $\code$ denote a linear $\nkdq$ code over $\GF{q}$ with block-length $n$, dimension $k$, and minimum distance $d$. 
For any codeword $\cw\in\code$, $\cw_i$ is said to be the $i$-th symbol of the codeword $\cw$. 

We say that the $i$-th symbol of a code $\code$ has locality $r$ if its value can be recovered from some other $r$ symbols of $\code$. The formal definition of locality is as follows (see~\cite{Gopalan:12}).

\begin{definition}
\label{def:locality}
[Locality] We say that the $i$-th coordinate of a code $\code$ has locality $r$ if there exists a set \mbox{$\rep{}{i}\subset[n]\setminus\{i\}$,} $|\rep{}{i}| \leq r$, such that, for every codeword $\cw\in\code$, \mbox{$\cw_i = \sum_{l\in\rep{}{i}}\lambda_{l}\cw_l$,} where $\lambda_l \in\GF{q}\setminus\{0\},$ $\forall\: l\in\rep{}{i}$. 
We say that $\rep{}{i}$ is a {\it repair group} of the $i$-th coordinate and define $\gam{i} = \{\cwi\cup\rep{}{i}\}$.

We say that an $\nkdq$ code has ({\it all-symbol}) {\it locality} $r$ if each of its $n$ coordinates has locality $r$. An LRC with these parameters is referred to as an $(n,k,r)$ LRC.
\end{definition}

Equivalently, we say that the coordinate $i$ has locality $r$, if the dual code $\dual$ contains a codeword $\cw'$ of Hamming weight at most $r+1$ such that the $i$-th coordinate is in the support of $\cw'$.



\begin{example}
\label{ex:LRC}
Let us consider a $(7,3)$ Simplex code $\code$, which is a dual of a $(7,4)$ Hamming code. In particular, $\code$ encodes three information symbols $\{a,b,c\}$ into seven symbols as $\{a,b,c,a+b,a+c,b+c,a+b+c\}$. It is easy to see that any symbol can be recovered from two other symbols. For instance, $a$ can be recovered from $b+c$ and $a+b+c$.\footnote{In fact, every symbol of the $(7,3)$ simplex code has three disjoint repair groups~\cite{CadambeM:15}. Further, note that, even though the $(7,3)$ simplex code is not optimal with respect to the distance upper bound in~\eqref{eq:d-upper-bound}, it is optimal with respect to a field size dependent rate upper bound established in~\cite{CadambeM:15}.}
\end{example}

In~\cite{Gopalan:12}, it is shown that the minimum distance $\dm{\code}$ of an $(n,k,r)$ LRC $\code$ is upper bounded as 
\begin{equation}
\label{eq:d-upper-bound}
\dm{\code} \leq n - k - \left\lceil\frac{k}{r}\right\rceil + 2.
\end{equation} 


Further, the authors of prove that any systematic code with locality for information symbols that achieves equality in~\eqref{eq:d-upper-bound} must follow a specific {\it structure}~\cite{Gopalan:12}. 
We state below the structure theorem \cite[Theorem 9]{Gopalan:12}, adapted to the form useful for our setup.
\begin{proposition} 
\label{prop:structure-theorem}
\cite{Gopalan:12} Let $\code$ be an $(n,k,r)$ code, where $r\mid k$, $r < k$, and $n = k + k/r$. Then, for any $i,j\in[n]$, $i\ne j$, we have either $\gam{i} = \gam{j}$ or $\gam{i}\cap\gam{j} = \emptyset$.
\end{proposition}

\subsection{Cooperative Locally Recoverable Codes}
\label{sec:cooperative-LRCs} 
Let $\code$ denote a linear $\nkdq$ code over $\GF{q}$ with block-length $n$, dimension $k$, and minimum distance $d$. 
We say that the code has $(r,\ell)$-cooperative locality if for every codeword, it is possible to repair any $\ell$ symbols from at most $r$ other symbols. The formal definition is as follows (see~\cite{Rawat:14:cooperative}).

\begin{definition}
\label{def:cooperative-locality}
We say that an $[n,k,d]$ code $\code$ has $(r,\ell)$-cooperative locality, if for any subset of $\ell$ {coordinates} $\Delta\subset[n]$, $|\Delta| = \ell$, there exists a set $\Gamma(\Delta)\subset[n]$ satisfying $\Delta\cap\Gamma(\Delta) = \emptyset$, $|\Gamma(\Delta)| \leq r$, such that, for every codeword $\cw\in\code$, the symbols $\cw_{\Delta}$ can be recovered using the {symbols} $\cw_{\Gamma(\Delta)}$.
\end{definition}
An LRC with these parameters is referred to as an $(n,k,r,\ell)$ cooperative LRC. Note that when $\ell = n-k$ and $r  = k$, then the above definition coincides with that of an MDS code. 

In~\cite{Rawat:15:cooperative}, it is shown that the minimum distance $\dm{\code}$ of an $(n,k,r,\ell)$ cooperative LRC $\code$ for $r\geq \ell$ is upper bounded as 
\begin{equation}
\label{eq:cooperative-d-upper-bound}
\dm{\code} \leq n - k +1 - \ell\left(\left\lceil\frac{k}{r}\right\rceil - 1\right).
\end{equation}


\section{Equivalence Results for Scalar-Linear Schemes}
\label{sec:duality-scalar-linear}

In this section, we consider non-interactive (single round), 
scalar-linear PIR-SI schemes.
In particular, for any given query $\Q{W}{S}$, the answer $\Ab{W}{S}$ can be specified as
\begin{equation}
    \label{eq:answer}
        \Ab{W}{S} = E\boldsymbol{X},
\end{equation}
where the matrix $E \in \GF{q}^{T\times K}$ depends on  $\Q{W}{S}$. We refer to $E$ as a {\it solution} to the PIR-SI problem. 
Note that $T$, the number of rows of $E$, denotes the number of symbols downloaded from the server.


\subsection{Single-Message PIR-SI Schemes and LRCs}
\label{sec:SM-PIR-SI-LRCs}

In this section, we show that a single-message PIR-SI scheme is {\it equivalent} to a locally recoverable code (LRC). In particular, we show that any solution to the single-message PIR-SI problem (SM-PIR-SI) must be a parity check matrix of an LRC. Furthermore, we show that it is possible to construct a solution to the SM-PIR-SI problem using a parity check matrix of an LRC.


First, we establish the relation from a solution of the SM-PIR-SI problem to a parity check matrix of an LRC. 
\begin{theorem}
\label{thm:1-PIR-LRC}
Any scalar-linear solution $E$ to the single-message PIR-SI problem must be a parity check matrix of an LRC with block length $K$ and locality $M$.
\end{theorem}
\begin{IEEEproof}
First, we note that the following necessary condition is imposed by the privacy and recoverability conditions. For any query $Q^{[W,S]}$, the answer $E$ should satisfy the following necessary condition: for any candidate demand index $W'\in [K]$, there must exist a potential side information index set $S'\subseteq [K]\setminus W'$, $|S'|\leq M$ such that it is possible to recover $W'$ from $EX$ and $X_{S'}$. In other words, the following condition must hold:
\begin{equation}
    \label{eq:necessary-cond-D-1}
    \evec{W'} \in \subspace{\twomatrix{E}{I_{S'}}}.
\end{equation} 
If the aforementioned necessary condition does not hold, then the server will learn from $E$ that $W'$ is not the user's demand index. {Indeed,} since $E$ is the solution corresponding to the query $Q^{[W,S]}$, we have
\begin{equation}
    \label{eq:privacy-counter}
    \mathbb{P}\left(\boldsymbol{W} = W' \mid \Qb{\boldsymbol{W}}{\boldsymbol{S}} = Q^{[W,S]} \right) = 0,
\end{equation}
which, in turn, implies that $I\left(\boldsymbol{W}; \Qb{\boldsymbol{W}}{\boldsymbol{S}}\right) > 0$. 
This violates the $W$-privacy condition~\eqref{eq:W-privacy}.

The above condition~\eqref{eq:necessary-cond-D-1} implies that for every $W'\in[K]$, $\subspace{E}$ must contain a vector $\vect{v}$ of Hamming weight at most $M+1$ such that $W'\in\supp{\vect{v}}$. {According to} Definition~\ref{def:locality},  $\subspace{E}^{\perp}$ is an LRC with block-length $K$ and all-symbol locality $M$.
\end{IEEEproof}

Theorem~\ref{thm:1-PIR-LRC} has the following two immediate implications. First, it allows us to construct a class of LRCs using solutions to the SM-PIR-SI problem. More specifically, given a solution $E$ to the SM-PIR-SI problem with $K$ messages and side information size $M$, one can easily obtain an LRC with block-length $K$ and locality $M$ as $\code = \subspace{E}^{\perp}$. 

Now, consider the Partition-and-Code scheme proposed in~\cite{KadheGHERS2017} for the SM-PIR-SI problem. Let $K ={ \alpha(M+1)+\beta}$ for some $\alpha > 0$ and $0 \leq \beta < M+1$. 
In the P\&C scheme, the user first randomly partitions the $K$ messages into $(\alpha+1)$ subsets, each of size at most $M+1$, such that one of the subsets is $W\cup S'$ for some $S'\subseteq S$. The user then asks the server to send the sum of messages in each subset, resulting in the download cost of $\alpha+1$ symbols. 

Note that the Partition-and-Code scheme yields a solution $E$ of size ${(\alpha+1)\times K}$ with the following form (up to column permutation): 
\begin{equation}
    \label{eq:PaC-E}
    E = \begin{bmatrix}
    \mathbf{1}_{M+1} & \mathbf{0}_{M+1} & \cdots & \mathbf{0}_{\beta}\\
    \mathbf{0}_{M+1} & \mathbf{1}_{M+1} & \cdots & \mathbf{0}_{\beta}\\
    \vdots & \vdots & \ddots & \vdots\\
    \mathbf{0}_{M+1} & \mathbf{0}_{M+1} & \cdots & \mathbf{1}_{\beta}
    \end{bmatrix},
\end{equation}
{It is easy to verify} that the corresponding LRC $\code = \subspace{E}^{\perp}$ is a direct-sum of $\alpha+1$ single-parity check codes, each of length at most $M+1$. In other words, $\code$ is a simple LRC that partitions the message symbols into $\alpha+1$ subsets each of size at most $M+1$, and adds a parity check symbol for each subset. 

Second, Theorem~\ref{thm:1-PIR-LRC} enables us to  use~\eqref{eq:d-upper-bound} to obtain an upper bound on the capacity of a (scalar-linear) single-message PIR-SI scheme. As we show next, the bound coincides with the upper bound derived in~\cite{KadheGHERS2017:Allerton,KadheGHERS2017}.

\begin{corollary}
    \label{cor:1-PIR-LRC}
    The scalar-linear capacity of the single-message PIR-SI problem is upper bounded by $\lceil K/(M+1) \rceil^{-1}$.
\end{corollary}
\begin{IEEEproof}
Let $E$ be a scalar-linear solution to the SM-PIR-SI problem. Let $\code = \subspace{E}^{\perp}$. Suppose the minimum distance of $\code$ is $d$. Note that we must have $d\geq 2$. For, if $d = 1$, $E$ must contain a column of all zeros. {Let $W'$ denote the index of this all-zero column. However, this implies that $X
_{W'}$ cannot be the demand, and this will violate the privacy.\footnote{Note that here we are using the same argument as in the proof of Theorem~\ref{thm:1-PIR-LRC} (cf.~\eqref{eq:privacy-counter}).}}
Now, since $\subspace{E}^{\perp}$ is an LRC with block-length $n = K$, dimension $k = K-T$, and locality $r = M$ from Theorem~\ref{thm:1-PIR-LRC}, we have from~\eqref{eq:d-upper-bound} that
$$K \geq K - T + \left\lceil\frac{K-T}{M} \right\rceil - 2 + d.$$
After re-arranging, and noting that $d\geq 2$ and $T$ is an integer, we get
$$T\geq \left\lceil\frac{K}{M+1}\right\rceil.$$
As the messages are independent and uniformly distributed over $\GF{q}$, we have $\Hp{\Ab{W}{S}} = T\log q$. The result then follows from~\eqref{eq:rate}.
\end{IEEEproof}

\begin{remark}
\label{rem:Tamo-Barg-rate-bound}
The above result can be directly proved using an upper bound on the rate of an LRC with locality $r$ given as $r/(r+1)$ (see~\cite[Theorem 1]{TamoB:14}). It is interesting to note that~\cite[Theorem 1]{TamoB:14} uses an argument based on acyclic induced subgraphs similar to~\cite{KadheGHERS2017:Allerton,KadheGHERS2017}.
\end{remark}

We say that a scalar-linear solution to SM-PIR-SI problem is an {\it optimal} solution, if $T = \lceil K/(M+1) \rceil$.
Then, Proposition~\ref{prop:structure-theorem} 
implies the following structure on any optimal scalar-linear solution. 
\begin{corollary}
\label{cor:structure-theorem-PIR}
{When $(M+1)\mid K$, any optimal scalar-linear solution $E$ to the PIR-SI problem can be converted  to the following form using elementary row operations and column permutations:}
\begin{equation}
    \label{eq:canonical-E}
    E = \begin{bmatrix}
    \times & \cdots & \times & 0 & \cdots & 0 & \cdots & 0 & \cdots & 0\\
    0 & \cdots & 0 & \times & \cdots & \times & \cdots & 0 & \cdots & 0\\
    \vdots & \ddots & \vdots & \vdots & \ddots & \vdots & \ddots & \vdots & \ddots & \vdots\\
    0 & \cdots & 0 & 0 & \cdots & 0 & \cdots & \times 
    & \cdots & \times
    \end{bmatrix},
\end{equation}
where $\times$ can be any non-zero element in $\GF{q}$, \ie, $\times\in\GF{q}\setminus\{0\}$, and the number of non-zero entries in each row is exactly $M+1$.
\end{corollary}
Since the solution obtained using the partition-and-code scheme (cf.~\eqref{eq:PaC-E}) has the same form as~\eqref{eq:canonical-E}, this shows the {\it uniqueness} of the solution obtained by the partition-and-code scheme. In other words, any optimal scalar-linear solution can be obtained from the partition-and-code solution using elementary row operations and column permutations. 

Next, we establish the relation from a parity check matrix of an LRC to a solution of the SM-PIR-SI problem.

\begin{theorem}
\label{thm:1-LRC-PIR}
Let $H$ be a parity check matrix of an LRC with block length $K$ and locality $M$. Then, it is possible to construct a single-message PIR-SI scheme, such that the solution $E$ is a column-permutation of $H$.
\end{theorem}
\begin{IEEEproof}
We present a constructive proof. In the rest of the proof, we consider all sets as ordered sets (with a natural ascending order). 
For a given $W$ and $S$, the user first finds a permutation $\pi$ on $[K]$ as follows. 
Choose an index $W'$ uniformly at random from $[K]$, independent of $W$ and $S$. 
Let $R(W')$ be a repair group of $W'$. If a coordinate has multiple repair groups, arbitrarily choose one repair group.\footnote{This arbitrary choice of a repair group for each coordinate is made {\it a priori}, and are known to the server as a part of the scheme.} By the definition of locality, we have $|R(W')| \leq M$. For simplicity, we assume that every repair group of any symbol is of size $M$.\footnote{The arguments can be easily generalized to the case when some repair groups are smaller than $M$.} Let $R'(W')$ be a random permutation of $R(W')$. 
Let $P = [K]\setminus\{W\cup S\}$, and $P'$ be a random permutation of $[K]\setminus\{W'\cup R(W')\}$. Let $\pi$ be the permutation that maps $W$ to $W'$, $S$ to $R'(W')$, and $P$ to $P'$. The user sends $\pi$ as its query $Q^{[W,S]}$. 
The server then applies $\pi$ to the columns of $H$ to obtain $E$, \ie, $E_{i} = H_{\pi(i)}$ for each $i\in[K]$, where $H_j$ is the $j$th column of $H$. Then, the server computes the answer as $EX$.


Next, we show that the above scheme satisfies the recoverablity and $W$-privacy conditions. Indeed, by the definition of locality for $W'$, $\subspace{H}$ contains a vector whose support is $W'\cup R(W')$. Therefore, by the construction of $E$, $\subspace{E}$ contains a vector whose support is $W\cup S$. Hence, the recoverability condition in~\eqref{eq:decodability} is satisfied. 

For the $W$-privacy, it suffices to show that, for any $W\in[K]$ and any permutation $\pi$, 
\begin{equation}
    \label{eq:privacy-step-1}
    \mathbb{P}\left(\Qb{\boldsymbol{W}}{\boldsymbol{S}} = \pi \mid \boldsymbol{W} = W\right) = \frac{1}{K!}.
\end{equation} 
This is because using~\eqref{eq:privacy-step-1}, it is easy to show that $\mathbb{P}\left(\boldsymbol{W} = W \mid \Qb{\boldsymbol{W}}{\boldsymbol{S}} = \pi\right) = \mathbb{P}\left(\boldsymbol{W} = W\right)$, from which the privacy condition~\eqref{eq:W-privacy} follows. 

Now, we give a proof of~\eqref{eq:privacy-step-1}. Observe that the query generation process first maps the demand index to a random index in $[K]$.
Let $\boldsymbol{W}'$ denote that  random index. 
Let $\boldsymbol{R}'(\boldsymbol{W}')$ and $\boldsymbol{P}'$ be random variables corresponding to (independent) uniform random permutations of $R(\boldsymbol{W}')$ and $[K]\setminus \{\boldsymbol{W}'\cup R(\boldsymbol{W}')\}$, respectively.   
Now, given a permutation $\pi$ on $[K]$ as a query, define the following events:
\begin{IEEEeqnarray}{rCl}
\IEEEeqnarraymulticol{3}{l}{\mathbb{P}\left(\Qb{\boldsymbol{W}}{\boldsymbol{S}} = \pi \mid \boldsymbol{W} = W\right)}\nonumber\\*
& \stackrel{(a)}{=} & \mathbb{P}\left(\boldsymbol{E}_1 | \boldsymbol{W} = W \right)\times
\mathbb{P}\left(\boldsymbol{E}_2 \mid \boldsymbol{E}_1, \boldsymbol{W} = W \right)\nonumber\\
& & \quad \times\: \mathbb{P}\left(\boldsymbol{E}_3 \mid \boldsymbol{E}_2, \boldsymbol{E}_1, \boldsymbol{W} = W \right),\nonumber\\
\qquad & \stackrel{(b)}{=} & \frac{1}{K}\times \frac{1}{M!\binom{K-1}{M}}\times \frac{1}{(K-1-M)!},\nonumber\\
\label{eq:indep-Q-W}
& = & \frac{1}{K!}\nonumber,
\end{IEEEeqnarray}
Then, for any $W\in[K]$ and a permutation $\pi$ on $[K]$, the probability of choosing $\pi$ as a query can be written as
\begin{IEEEeqnarray}{rCl}
\IEEEeqnarraymulticol{3}{l}{\mathbb{P}\left(\Qb{\boldsymbol{W}}{\boldsymbol{S}} = \pi \mid \boldsymbol{W} = W\right)}\nonumber\\*
& \stackrel{(a)}{=} & \mathbb{P}\left(\boldsymbol{E}_1 | \boldsymbol{W} = W \right)\times
\mathbb{P}\left(\boldsymbol{E}_2 \mid \boldsymbol{E}_1, \boldsymbol{W} = W \right)\nonumber\\
& & \quad \times\: \mathbb{P}\left(\boldsymbol{E}_3 \mid \boldsymbol{E}_2, \boldsymbol{E}_1, \boldsymbol{W} = W \right),\nonumber\\
\qquad & \stackrel{(b)}{=} & \frac{1}{K}\times \frac{1}{M!\binom{K-1}{M}}\times \frac{1}{(K-1-M)!},\nonumber\\
\label{eq:indep-Q-W}
& = & \frac{1}{K!}\nonumber,
\end{IEEEeqnarray}
where (a) follows from the query generation procedure, and (b) uses~\eqref{eq:SideInfoDist} and~\eqref{eq:WantSetDist} to compute $\mathbb{P}\left(\boldsymbol{E}_2 \mid \boldsymbol{E}_1, \boldsymbol{W} = W \right)$. This completes the proof of~\eqref{eq:privacy-step-1}, and concludes the proof.
\end{IEEEproof}


\subsection{Multi-Message PIR-SI and Cooperative LRCs}
\label{sec:duality-multi-message-main}

In this section, we show that a multi-message PIR-SI scheme is a dual of a cooperative LRC, 
introduced in~\cite{Rawat:14:cooperative}. 


First, we show that any solution to the multi-message PIR-SI problem should be a parity check matrix of a code with cooperative locality.
\begin{theorem}
\label{thm:D-PIR-LRC}
Any scalar-linear solution $E$ to the multi-message PIR-SI problem {with a demand set of size $D$ and a side information set of size $M$ } must be a parity check matrix of an LRC with block length $K$ and $(M,D)$-cooperative locality.
\end{theorem}
\begin{IEEEproof}
First, we note that the following necessary condition is imposed by the privacy and recoverability conditions. For any query $Q^{[W,S]}$, the answer $E$ should 
{satisfy the following necessary condition:}
for every candidate demand index set $W'\in [K]$, $|W'| = D$, there must exist a potential side information index set $S'\subseteq [K]\setminus W'$, $|S'|\leq M$ such that it is possible to recover $X_{W'}$ from $EX$ and $X_{S'}$. In other words, {the following condition must hold:}
\begin{equation}
    \label{eq:necessary-cond-D-D}
    \evec{i_j} \in \subspace{\twomatrix{E}{I_{S'}}}, \quad \forall\: i_j\in W'.
\end{equation} 
If the aforementioned necessary condition does not hold, then the server will learn from $E$ that $W'$ is not the user's demand index. Since $E$ is the solution corresponding to the query $Q^{[W,S]}$, we have
$$\mathbb{P}\left(\boldsymbol{W} = W' \mid \Qb{\boldsymbol{W}}{\boldsymbol{S}} = Q^{[W,S]} \right) = 0,$$
which, in turn, implies that $I\left(\boldsymbol{W}; \Qb{\boldsymbol{W}}{\boldsymbol{S}}\right) > 0$. This violates the $W$-privacy condition~\eqref{eq:W-privacy}.  This violates the privacy condition~\eqref{eq:W-privacy}. 

The above condition~\eqref{eq:necessary-cond-D-D} implies that for every subset  $W'=\{i_1,i_2,\ldots,i_D\}\subseteq[K]$ of size $D$, $\subspace{E}$ must contain $D$ vectors $v_{1},v_{2},\ldots,v_{D}$ such that $|\cup_{j=1}^{D}\supp{v_j}|\leq D+M$, and for each $1\leq j\leq D$, $\supp{v_j}\cap W' = \{i_j\}$. It is {easy} 
to verify from Definition~\ref{def:cooperative-locality} that $\subspace{E}^{\perp}$ is an $(M,D)$ cooperative LRC with block-length $K$.
\end{IEEEproof}

\begin{corollary}
    \label{cor:D-PIR-LRC}
    For $M\geq D$, the scalar-linear capacity of the multi-message PIR-SI problem is upper bounded by ${D}/{\lceil DK/(M+D) \rceil}$. 
\end{corollary}
\begin{IEEEproof}
Let $\code = \subspace{E}^{\perp}$. Note that from Theorem~\ref{thm:D-PIR-LRC}, $\code$ must be a code with blocklength $K$ and $(M,D)$-cooperative locality. Using~\eqref{eq:cooperative-d-upper-bound}, it is shown in~\cite[Corollary 1]{Rawat:15:cooperative} that the rate of a code with $(M,D)$-cooperative locality for $M\geq D$ is upper bounded as $M/(M+D)$. 
Therefore, we have $T/K \geq 1 - M/(M+D)$. This yields $T \geq \lceil DK/(D+M) \rceil$,
which gives the capacity upper bound.
\end{IEEEproof}

Next, we show that it is possible to construct a solution to the multi-message PIR-SI problem using a parity check matrix of a cooperative locality code.

\begin{theorem}
\label{thm:D-LRC-PIR}
Let $H$ be a parity check matrix of an LRC with block-length $K$ and $(D,M)$-cooperative locality. Then, it is possible to construct a multi-message PIR-SI scheme, such that the solution $E$ is a column-permutation of $H$.
\end{theorem}
\begin{IEEEproof}
The query generation process and the rest of the proof is similar to the proof of Theorem~\ref{thm:1-PIR-LRC}.
\end{IEEEproof}

\begin{corollary}
    \label{cor:rate-LRC}
    For $\ell > r$, the rate of a linear $(n,k,r,\ell)$ cooperative LRC is upper bounded by $r/n$.
\end{corollary}
\begin{IEEEproof}
Let $H$ be a parity check matrix of an $(n,k,r,\ell)$ cooperative LRC. From Theorem~\ref{thm:D-PIR-LRC}, $H$ is a solution (up to a column-permutation) of a multi-message PIR-SI problem such that $K = n$, $M = r$, and $D = \ell$. Now, in~\cite[Lemma 1]{HeidarzadehGKES:18}, it is shown that, when $D > M$, the number of transmissions in any multi-message PIR-SI scheme is at least $K - M$. Therefore, we have $n - k \geq n - r$, from which the result follows.
\end{IEEEproof}

\begin{remark}
\label{rem:rate-LRC}
Corollary~\ref{cor:rate-LRC} yields a better bound on the rate of a cooperative LRC for $\ell > r$ than~\cite[Corollary 1]{Rawat:15:cooperative} given as $r/(r+\ell) + \ell^2/(nr)$. In fact, the rate bound is tight for $n > 2r$. This is because an $(n,r)$ MDS code trivially has $(r,\ell)$-cooperative locality for any $\ell\geq r$.
\end{remark}

Theorem~\ref{thm:D-PIR-LRC} also enables us to obtain {\it computationally efficient} multi-message PIR-SI solutions. In particular, for ${D\leq M}$, the schemes in~\cite{HeidarzadehGKES:18} (see also~\cite{LiGastpar:18}) rely on generalized Reed-Solomon codes, and thus, require a finite field size at least $M + \lceil M/D \rceil$. 
On the other hand, it is possible to use constructions of cooperative LRCs to obtain PIR-SI schemes over smaller field size.\footnote{Note that small field size schemes obtained from cooperative LRCs may have smaller download rate than those in~\cite{HeidarzadehGKES:18,LiGastpar:18}.} As an example, an $(n = 2^k-1,k)$ simplex code has $(\ell+1,\ell)$-cooperative locality for any $1\leq \ell \leq (n-1)/2$ (see~\cite{Rawat:15:cooperative}). Thus, it is possible to obtain multi-message PIR-SI solutions over the binary field when $K = 2^t - 1$ for a positive integer $t$, $1\leq D\leq (K-1)/2$, and $M = D+1$.

\subsection{$(W,S)$-Private PIR-SI Schemes and MDS Codes}
\label{sec:duality-WS-private}

In this section, we show an equivalence between a solution to the $(W,S)$-PIR-SI problem and a maximum distance separable (MDS) code. 

First, we establish the relation from a solution of the $(W,S)$-PIR-SI problem to a parity check matrix of an MDS code.
\begin{theorem}
\label{thm:PIR-MDS}
Any scalar-linear solution $E$ to the $(W,S)$-PIR-SI problem must be a parity check matrix of a $(K,M)$ MDS code.
\end{theorem}
\begin{IEEEproof}
First, we note that the $(W,S)$-privacy condition implies the following necessary condition: for each message $X_i$ and every set $S_i\subseteq[K]\setminus\{i\}$ of size $M$, it is possible to recover $X_i$ from $EX$ and $X_{S_i}$. If this is not the case, then the server learns that the user cannot possess $X_{S_i}$ and demand any $X_W$ such that $i\in W$. {Indeed,} since $E$ is the solution corresponding to the query $Q^{[W,S]}$, we have
\begin{equation}
    \label{eq:privacy-counter}
    \mathbb{P}\left(\boldsymbol{S} = S_i, i\in\boldsymbol{W} \mid \Qb{\boldsymbol{W}}{\boldsymbol{S}} = Q^{[W,S]} \right) = 0,
\end{equation}
which, in turn, implies that $I\left(\boldsymbol{W},\boldsymbol{S}; \Qb{\boldsymbol{W}}{\boldsymbol{S}}\right) > 0$. 
This violates the $(W,S)$-privacy condition~\eqref{eq:WS-privacy}.

The aforementioned necessary condition implies that, for any set $S\subset[K]$ of size $M$, for every $i\in[K]\setminus S$, we should have
\begin{equation}
    \label{eq:necessary-cond-WS}
    \evec{i} \in \subspace{\twomatrix{E}{I_{S}}}.
\end{equation} 
Equation~\eqref{eq:necessary-cond-WS}, in turn, implies that the columns of $E$ in $[K]\setminus S$ must be linearly independent. Since this should hold for each subset $S\subset[K]$ of size $M$, we have that every subset of columns of $E$ of size $K-M$ are linearly independent. Thus, $E$ must be a parity check matrix of a $(K,M)$ MDS code.
\end{IEEEproof}

Next, we establish a relation from a parity check matrix of an MDS code to a solution of the $(W,S)$-PIR-SI problem. It is worth noting that the achievability schemes in~\cite{KadheGHERS2017,HeidarzadehGKES:18} for $(W,S)$-privacy are based on MDS codes.
\begin{theorem}
\label{thm:MDS-PIR}
Let $H$ be a parity check matrix of a $(K,M)$-MDS code. Then, $E = H$ is a solution to the $(W,S)$-PIR-SI problem.
\end{theorem}
\begin{IEEEproof}
First, note that the scheme with $E = H$ is private, since the solution is independent of the particular realization of $\boldsymbol{W}$ and $\boldsymbol{S}$. As the server already knows the size of the side information index set, it does not get any other information about $\boldsymbol{W}$ and $\boldsymbol{S}$ from $E$.

To see the recoverability, note that any $K-M$ columns of $H$ are linearly independent. Thus, given the side information $X_S$ for any $S\subset[K]$ of size $M$, the user can recover all the messages $X_i$, $i\in [K]\setminus S$, including the demand message(s) $X_W$.
\end{IEEEproof}

\section{Equivalence Results for Non-Linear Schemes}
\label{sec:duality-non-linear}

In this section, we consider generic PIR-SI schemes and LRCs, which encompass scalar-linear, vector-linear, and non-linear schemes. 
We begin with the definition of a generic LRC. 

\begin{definition}
\label{def:LRC-gen}
An $(n,k,r)$ LRC $\code\subseteq\GF{q}^n$ is a set of vectors in $\GF{q}^n$ of size $q^k$, {referred to as \emph{codewords},} together with 
\begin{enumerate}
    \item an encoding function $f:\GF{q}^k \rightarrow \code$, which is a bijection between vectors in $\GF{q}^k$ and codewords in $\code$, and
    \item a set of deterministic repair functions $g_1, g_2, \ldots, g_n$, \mbox{$g_i:\GF{q}^{r} \rightarrow \GF{q}$,} such that, for every coordinate $i\in[n]$, there exists a set of coordinates $R(i)\subset[n]\setminus\{i\}$, $|R(i)| = r$ satisfying $g_i(\cw_{R(i)}) = \cwi$ for every codeword $\cw\in\code$. We say that $R(i)$ is a repair group of the $i$-th coordinate. 
\end{enumerate}
\end{definition}

Next, for the SM-PIR-SI problem, we define a PIR-SI code. Towards this end, we introduce the following notation:
\begin{equation}
    \label{eq:class-WS}
    \mathcal{W} = \left\{(W,S) \mid W\in[K], S\subset[K]\setminus\{W\}, |S| = M \right\}.
\end{equation}
{That is, $\mathcal{W}$ is the set of all possible combinations of the demand index and the side information index set.}

\begin{definition}
\label{def:PIR-gen}
A PIR-SI code for $\GF{q}^K$ is a set of {vectors} in $\GF{q}^T$, {referred to as \emph{codewords},}  together with
\begin{enumerate}
    \item  a class of deterministic answer functions $\setA$, where each function $A\in\setA$ maps vectors from $\GF{q}^K$ to the codewords, i.e., {$A:\GF{q}^K\rightarrow  \GF{q}^T$,}
    \item a class of deterministic recovery functions $\mathcal{D}$, where each function \mbox{$D\in\mathcal{D}$} is from $\GF{q}^{T+M}$ to $\GF{q}$, and
    \item a stochastic query function $Q:\mathcal{W}\rightarrow\setA$ that maps $(W,S)$ to an answer function $A\in\setA$ (independently of {the value of} $X_S$) such that:
    \begin{itemize}
        \item[(i)] for every $W',W\in[K]$, $S\subset[K]\setminus\{W\}$, $|S| = M$, {and for each $A\in\setA$}, 
    \begin{equation}
        \label{eq:W-privacy-gen}
        \mathbb{P}\left(\boldsymbol{W} = W'\mid Q(W,S) = A\right) = \mathbb{P}\left(\boldsymbol{W} = W'\right),
    \end{equation}
    and 
    \item[(ii)] there exists a decoding function $D\in\mathcal{D}$ satisfying
    \begin{equation}
        \label{eq:decodability-gen}
        D\left(A(X_1,\cdots,X_K),X_S\right) = X_W.
    \end{equation} 
    \end{itemize}
\end{enumerate}
We refer to $T$ as the length of the PIR code.

\end{definition}

It is straightforward to show that the $W$-privacy condition~\eqref{eq:W-privacy-gen} implies the following necessary condition on a PIR code.
\begin{lemma}
\label{lem:necessary-cond} 
In a PIR-SI code, for any $A\in\setA$, for every $j\in[K]$, there must exist a decoding function $D_j\in\mathcal{D}$ and a set  \mbox{$S_j\subset[K]\setminus\{j\}$,} $|S_j| \: = M$, such that $D_j\left(A(X_1,\cdots,X_K),X_{S_j}\right) = X_j$. 
\end{lemma}

Now, we show a relation from a PIR-SI code to an LRC. It is worth noting that the proof technique is similar to~\cite[Lemma 3]{Mazumdar:15:capacity}.
\begin{theorem}
\label{thm:1-PIR-LRC-gen}
Given a PIR-SI code of length $T$ over $\GF{q}$, it is possible to construct an LRC of size (at least) $q^{K-T}$.
\end{theorem}
\begin{IEEEproof}
First, note that, for any $A\in\setA$, there must exist a vector $\vect{a}\in\GF{q}^T$ such that $\left|\left\{X\in\GF{q}^K \mid A(X) = \vect{a} \right\}\right| \geq q^{K-T}$. This is because every $A\in\setA$ maps $\GF{q}^K$ to $\GF{q}^T$.
Next, for an arbitrary $A\in\setA$ and the corresponding $\vect{a}$, let us define $\code_{\vect{a}} = \left\{X\in\GF{q}^K \mid A(X) = \vect{a} \right\}$. 
Now, from Lemma~\ref{lem:necessary-cond}, for every $i\in[K]$, there must exist a deterministic decoding function $D_i$ and a set  $S_i\subset[K]\setminus\{i\}$, $|S_i| = M$, such that $D_i\left(\vect{a},X_{S_i}\right) = X_i$. Using this, define, for every $i\in[K]$, $R(i) = S_i$, and $g_i\left(\cw_{R(i)}\right) = D_i\left(\vect{a},X_{S_i}\right)$. It is easy to verify that the set $\code_{\vect{a}}$ along with with an arbitrary bijection $E:\GF{q}^{\lfloor \log_q|\code_{\vect{a}}|\rfloor}\rightarrow\code$ and repair functions $g_1, g_2, \ldots, g_K$ is an LRC of size at least $q^{K-T}$.
\end{IEEEproof}

Next, from~\cite[Theorem 2.1]{TamoB:14}, we have the following upper bound on the size of an $(n,k,r)$ LRC.
\begin{proposition}
\label{prop:max-size} 
\cite{TamoB:14} For any $(n,k,r)$ LRC $\code\subset\GF{q}^n$, the size $|\code| \leq q^{n - \lceil n/(
r+1) \rceil}$. 
\end{proposition}
We refer to an $(n,k,r)$ LRC $\code$ satisfying the equality $|\code| = q^{n - \lceil n/(
r+1) \rceil}$ to be an {\it optimal} LRC.

To complete the equivalence, we establish a relation from an optimal LRC to a PIR-SI code.
\begin{theorem}
\label{thm:1-LRC-PIR-gen}
Given an optimal $(K,K - \lceil K/(M+1)\rceil,M)$ LRC, 
it is possible to construct a PIR-SI code of length \mbox{$\lceil K/(M+1) \rceil$} over $\GF{q}$.
\end{theorem}

In order to prove Theorem~\ref{thm:1-LRC-PIR-gen}, we need two other lemmas.  
{To simplify the presentation, we define} $T_{OPT} \triangleq \lceil K/(M+1) \rceil$. Also, for a code $\code$ of block-length $K$ and a set $P\subset[K]$, let $\code_P$ denote the code obtained by puncturing $\code$ on the coordinates outside of $P$. 

First, we show that any optimal LRC must contain $K - T_{OPT}$ coordinates such that values on these coordinates determine the values of the remaining $T_{OPT}$ coordinates. 
Note that for an arbitrary $(n, k)$ non-linear code, there my not exist any subset of $k$ coordinates that determine values of the remaining coordinates.

\begin{lemma}
\label{lem:good-coordinates}
For an optimal $(K,K - T_{OPT},M)$ LRC $\code$, there exists a partition of $K$ coordinates into sets $P_1$ and $P_2$ such that {$|P_1| = K - T_{OPT}$,} $|P_2| = T_{OPT}$, and for any codeword $\cw\in\code$, the symbols $\cw_{P_2}$ can be recovered from the symbols $\cw_{P_1}$.
\end{lemma}
\begin{IEEEproof}
We iteratively construct $P_1$ and $P_2$ as follows. 
\begin{itemize}
    \item[1.] Initialize $P_1 = P_2 = \emptyset$
    \item[2.] {\it While} $|P_1 \cup P_2| < K$:
    \begin{itemize}
        \item[2.1] Choose a coordinate $i \not\in P_1 \cup P_2$
        \item[2.2] Set {$P_1 \leftarrow P_1 \cup R(i)$}, for a repair group $R(i)$ of $i$
        \item[2.3] Set {$P_2 \leftarrow P_2 \cup \{i\}$.}
    \end{itemize}
\end{itemize}
By the construction of $P_1$ and $P_2$, the coordinates in $P_2$ can be recovered from the coordinates in $P_1$. 


Note that, in each step, $P_2$ grows by one, and $P_1$ grows by at most $M$ as the locality of the code is $M$. In other words, in each step, $P_1\cup P_2$ grows by at most $M+1$. Therefore, the number of steps for which the while loop runs is at least $\lceil K/(M+1) \rceil = T_{OPT}$. This gives $|P_2| \geq T_{OPT}$.

Next, we show that $|P_2| \leq  T_{OPT}$. Since there is a bijection between $\GF{q}^{K-T_{OPT}}$ and $\code$, and since the coordinates in $P_2$ are a function of those in $P_1$, there must be a bijection between $\GF{q}^{K-T_{OPT}}$ and $\code_{P_1}$. This implies that $|P_1| \geq K - T_{OPT}$, and thus, $|P_2| \leq T_{OPT}$.

{We conclude that $|P_2| = T_{OPT}$, which completes the proof}.
\end{IEEEproof} 

Given a vector $\vect{u}$, we define a translation of an LRC $\code$ as 
\begin{equation}
    \label{eq:translation}
    \code + \vect{u} = \left\{\cw + \vect{u} \mid \cw\in\code \right\}.
\end{equation}
Now, using Lemma~\ref{lem:good-coordinates}, we show that there exist $q^{T_{OPT}}$ translations of an optimal LRC that partition $\GF{q}^K$. 

\begin{lemma}
\label{lem:translations}
For an optimal $(K,K - \lceil K/(M+1)\rceil,M)$ LRC $\code$, there exist $q^{T_{OPT}}$ distinct vectors $\vect{u}_j\in\GF{q}^K$, $j = 0,\ldots,q^{T_{OPT}}-1$, such that the translations $\left\{\code + \vect{u}_j \mid j = 0,\ldots,q^{T_{OPT}}-1\right\}$ partition the space $\GF{q}^K$. That is, 
\begin{equation}
    \label{eq:translations-disjoint}
    (\code + \vect{u}_i) \cap (\code + \vect{u}_j) = \emptyset, \quad \forall \: i\ne j,
\end{equation} and
\begin{equation}
\label{eq:translations-cover}
\cup_{j=0}^{q^{T_{OPT}}-1}\left(\code + \vect{u}_j \right) = \GF{q}^K.
\end{equation}
\end{lemma}
\begin{IEEEproof}
We give a constructive proof. Let $P_1$ and $P_2$ be the sets of coordinates of $\code$ as described in Lemma~\ref{lem:good-coordinates}. Without loss of generality, let $P_1$ be the first $K - T_{OPT}$ coordinates. Let $\left\{\vect{v}_i \mid 0\leq i\leq q^{T_{OPT}}-1\right\}$ denote the set of vectors in $\GF{q}^{T_{OPT}}$ in a lexicographic order. For each $0\leq i\leq q^{T_{OPT}}-1$, define $\vect{u}_i = \left[\vect{0} \:\: \vect{v}_i\right]$, where $\vect{0}$ is the all-zero vector of length $K-T^{*}$. 

Note that any translation of $|\code$ has the same size as $\code$. Thus, to prove~\eqref{eq:translations-cover}, it suffices to show~\eqref{eq:translations-disjoint}.
We prove this by the way of contradiction. Suppose, for contradiction, that there exists a pair of codewords $\cw, \cw'\in\code$ such that $\cw + \vect{u}_i = \cw' + \vect{u}_j$. 
This implies that 
\begin{equation}
    \label{eq:contradiction}
    [\cw_{P_1} \:\: \cw_{P_2} + \vect{v}_i] = [\cw'_{P_1} \:\: \cw'_{P_2} + \vect{v}_j].
\end{equation}
Therefore, $\cw_{P_1} = \cw'_{P_1}$. Further, since the coordinates in $P_2$ can be recovered from those in $P_1$ (Lemma~\ref{lem:good-coordinates}), we must have $\cw_{P_2} = \cw'_{P_2}$. However, as $\vect{v}_i \ne \vect{v}_j$, we have a contradiction to~\eqref{eq:contradiction}.
\end{IEEEproof}


{\it Proof of Theorem~\ref{thm:1-LRC-PIR-gen}:}  
Lemma~\ref{lem:translations} enables us to construct a PIR-SI code of length $T_{OPT}$ over $\GF{q}$ using an optimal LRC $\code$ as follows. 

{\it Answer functions:} We construct a set $\setA$ of $K!$ answer functions, and associate every answer function with a permutation on $[K]$. 
Towards this end, we need the following additional notation. For $0\leq a\leq q^{T_{OPT}}-1$, let $\bar{a}_q$ denote the length-$T_{OPT}$ $q$-ary expansion of $a$. For a permutation $\pi$ on $[K]$ and a vector $[X_1 \cdots X_K] \in \GF{q}^K$, let $\pi(X) = X_{\pi([K])}$

Let $\mathcal{U} = \{\vect{u}_j\in\GF{q}^K, j = 0,\ldots,q^{T_{OPT}}-1\}$ be a set of vectors as described in Lemma~\ref{lem:translations}.  
For a given $X\in\GF{q}^K$ and a permutation $\pi$ on $[K]$, let $0\leq a\leq q^{T_{OPT}}-1$ be such that $\pi(X) \in \code + \vect{u}_a$. 
Note that, by Lemma~\ref{lem:translations}, the translations $\left\{\code + \vect{u}_j \mid 0\leq j\leq q^{T_{OPT}}-1\right\}$ partition the space $\GF{q}^K$. Hence, there exists a unique such $\vect{u}_a\in\mathcal{U}$ for every $X\in\GF{q}^K$ and any permutation $\pi$ on $[K]$.
Define the answer functions for every $X\in\GF{q}^K$ and every permutation $\pi$ on $[K]$ as
\begin{equation}
    \label{eq:encoder-PIR}
    A_{\pi}\left(X\right) = \bar{a}_q.
\end{equation}

{\it Query function:} We are given an index $W\in[K]$ and a set $S\subset[K]\setminus\{W\}$. First, choose an index $W'\in[K]$ uniformly at random independent of $W$ and $S$. Choose an arbitrary repair group of $W'$, say $R(W')$.\footnote{If a coordinate has multiple repair groups, arbitrarily choose one repair group. This arbitrary choice of a repair group for each coordinate is made {\it a priori}, and are known to the server as a part of the scheme.}
Let $P = [K]\setminus(W\cup S)$. Let $R'(W')$ and $P'$ be random permutations of sets $R(W')$ and $[K]\setminus(W'\cup R(W'))$, respectively. 
Let $\pi$ be a permutation on the set $[K]$ that maps $W$ to $W'$, $S$ to $R'(W')$, and $P$ to $P'$. Then, the query function $Q$ maps $(W,S)$ to $A_{\pi}$ in $\setA$. Note that it suffices for the user to send $\pi$ as their query.

{\it Recovery functions:} For a set $P\subset[K]$, let $\vect{u}_a\hspace{-1.1mm}\mid_P$ denote the length-$|P|$ vector obtained by deleting the coordinates of $\vect{w}_a$ outside $P$. 
Now, given $\pi$ and $A_{\pi}$, define the recovery function as
\begin{equation}
    \label{eq:decoding-PIR}
    D\left(A_{\pi}(X),X_S\right) = g_{W'}\left(X_{R'(W')}-\vect{u}_a\hspace{-1.1mm}\mid_{R'(W')}\right) + \vect{u}_a\hspace{-1.1mm}\mid_{W'},
\end{equation}
where $g_{W'}(\cdot)$ is the repair function of $\code$ for the coordinate $\cw_{W'}$ (see Definition~\ref{def:LRC-gen}).

{\it Recoverability and Privacy:} It is straightforward to verify that $D\left(A_{\pi}(X),X_S\right) = X_{W}$ (cf.~\eqref{eq:decoding-PIR}). The $W$-privacy condition~\eqref{eq:W-privacy-gen} can be proven in the same way as in the proof of Theorem~\ref{thm:1-LRC-PIR}, and thus, the proof is omitted.

\section{Conclusion}
\label{sec:conclusion}
The theoretical computer science community has established a strong relationship between PIR schemes and locally decodable codes. This paper extends this theme by establishing strong relationship between PIR schemes for a recently proposed single-server PIR with side information problem and locally recoverable codes. 
As corollaries to these results, we obtain upper bounds on the download rate for PIR-SI schemes, and a novel rate upper bound on cooperative LRCs. 

\section*{Acknowledgement}
S. Kadhe would like to thank Kannan Ramchandran for helpful discussions.





\bibliographystyle{IEEEtran}
\bibliography{PIR_salim,coding1,coding2,pir_bib}

\begin{thebibliography}{10}
\providecommand{\url}[1]{#1}
\csname url@samestyle\endcsname
\providecommand{\newblock}{\relax}
\providecommand{\bibinfo}[2]{#2}
\providecommand{\BIBentrySTDinterwordspacing}{\spaceskip=0pt\relax}
\providecommand{\BIBentryALTinterwordstretchfactor}{4}
\providecommand{\BIBentryALTinterwordspacing}{\spaceskip=\fontdimen2\font plus
\BIBentryALTinterwordstretchfactor\fontdimen3\font minus
  \fontdimen4\font\relax}
\providecommand{\BIBforeignlanguage}[2]{{%
\expandafter\ifx\csname l@#1\endcsname\relax
\typeout{** WARNING: IEEEtran.bst: No hyphenation pattern has been}%
\typeout{** loaded for the language `#1'. Using the pattern for}%
\typeout{** the default language instead.}%
\else
\language=\csname l@#1\endcsname
\fi
#2}}
\providecommand{\BIBdecl}{\relax}
\BIBdecl

\bibitem{chor1998private}
B.~Chor, E.~Kushilevitz, O.~Goldreich, and M.~Sudan, ``Private information
  retrieval,'' \emph{Journal of the ACM}, vol.~45, no.~6, pp. 965--981, 1998.

\bibitem{yekhanin2010private}
S.~Yekhanin, ``Private information retrieval,'' \emph{Communications of the
  ACM}, vol.~53, no.~4, pp. 68--73, 2010.

\bibitem{Sun2017}
\BIBentryALTinterwordspacing
H.~Sun and S.~A. Jafar, ``The capacity of private information retrieval,''
  \emph{CoRR}, vol. abs/1602.09134, 2016. [Online]. Available:
  \url{http://arxiv.org/abs/1602.09134}
\BIBentrySTDinterwordspacing

\bibitem{JafarPIR3new}
------, ``The capacity of robust private information retrieval with colluding
  databases,'' \emph{IEEE Trans. on Info. Theory}, vol.~64, no.~4, pp.
  2361--2370, April 2018.

\bibitem{tajeddine2017private1}
R.~Tajeddine and S.~El~Rouayheb, ``Robust private information retrieval on
  coded data,'' in \emph{2017 IEEE International Symposium on Information
  Theory (ISIT)}.\hskip 1em plus 0.5em minus 0.4em\relax IEEE, 2017.

\bibitem{Banawan2017}
\BIBentryALTinterwordspacing
K.~Banawan and S.~Ulukus, ``Multi-message private information retrieval:
  Capacity results and near-optimal schemes,'' \emph{CoRR}, vol.
  abs/1702.01739, 2017. [Online]. Available:
  \url{http://arxiv.org/abs/1702.01739}
\BIBentrySTDinterwordspacing

\bibitem{BU18}
------, ``The capacity of private information retrieval from coded databases,''
  \emph{IEEE Trans. on Info. Theory}, vol.~64, no.~3, pp. 1945--1956, March
  2018.

\bibitem{KadheGHERS2017:Allerton}
S.~Kadhe, B.~Garcia, A.~Heidarzadeh, S.~E. Rouayheb, and A.~Sprintson,
  ``Private information retrieval with side information: The single server
  case,'' in \emph{2017 55th Annual Allerton Conference on Communication,
  Control, and Computing (Allerton)}, Oct 2017, pp. 1099--1106.

\bibitem{KadheGHERS2017}
\BIBentryALTinterwordspacing
------, ``Private information retrieval with side information,'' \emph{CoRR},
  vol. abs/1709.00112, 2017. [Online]. Available:
  \url{http://arxiv.org/abs/1709.00112}
\BIBentrySTDinterwordspacing

\bibitem{HKGRS:2018}
A.~Heidarzadeh, B.~Garcia, S.~Kadhe, S.~E. Rouayheb, and A.~Sprintson, ``On the
  capacity of single-server multi-message private information retrieval with
  side information,'' in \emph{2018 56th Annual Allerton Conf. on Commun.,
  Control, and Computing}, Oct 2018.

\bibitem{LG:2018}
S.~Li and M.~Gastpar, ``Single-server multi-message private information
  retrieval with side information,'' in \emph{2018 56th Annual Allerton Conf.
  on Commun., Control, and Computing}, Oct 2018.

\bibitem{yekhanin2011locally}
S.~Yekhanin, ``Locally decodable codes,'' in \emph{Computer Science--Theory and
  Applications}.\hskip 1em plus 0.5em minus 0.4em\relax Springer, 2011, pp.
  289--290.

\bibitem{Gopalan:12}
P.~Gopalan, C.~Huang, H.~Simitci, and S.~Yekhanin, ``On the locality of
  codeword symbols,'' \emph{Information Theory, IEEE Transactions on}, vol.~58,
  no.~11, pp. 6925--6934, Nov 2012.

\bibitem{Rawat:14:cooperative}
A.~S. Rawat, A.~Mazumdar, and S.~Vishwanath, ``On cooperative local repair in
  distributed storage,'' in \emph{2014 48th Annual Conference on Information
  Sciences and Systems (CISS)}, March 2014, pp. 1--5.

\bibitem{Rawat:15:cooperative}
------, ``Cooperative local repair in distributed storage,'' \emph{Journal on
  Advances in Signal Processing}, vol. 2015, no.~1, p. 107, Dec 2015.

\bibitem{HeidarzadehGKES:18}
A.~Heidarzadeh, B.~Garcia, S.~Kadhe, S.~E. Rouayheb, and A.~Sprintson, ``On the
  capacity of single-server multi-message private information retrieval with
  side information,'' \emph{CoRR}, vol. abs/1807.09908, 2018.

\bibitem{CadambeM:15}
V.~R. Cadambe and A.~Mazumdar, ``Bounds on the size of locally recoverable
  codes,'' \emph{IEEE Transactions on Information Theory}, vol.~61, no.~11, pp.
  5787--5794, Nov 2015.

\bibitem{TamoB:14}
I.~Tamo and A.~Barg, ``A family of optimal locally recoverable codes,''
  \emph{Information Theory, IEEE Transactions on}, vol.~60, no.~8, pp.
  4661--4676, Aug 2014.

\bibitem{LiGastpar:18}
\BIBentryALTinterwordspacing
S.~Li and M.~Gastpar, ``Single-server multi-message private information
  retrieval with side information,'' \emph{CoRR}, vol. abs/1808.05797, 2018.
  [Online]. Available: \url{http://arxiv.org/abs/1808.05797}
\BIBentrySTDinterwordspacing

\bibitem{Mazumdar:15:capacity}
A.~{Mazumdar}, ``Storage capacity of repairable networks,'' \emph{IEEE Trans.
  on Info. Theory}, vol.~61, no.~11, pp. 5810--5821, Nov 2015.

\end{thebibliography}

\end{document}